\documentclass[aps,preprint,nofootinbib]{revtex4}
\usepackage{graphicx}
\begin{document}
\title{\Large \bf Superradiance and instability of small rotating  charged AdS  black holes in all dimensions}
\author{\large Alikram N. Aliev}
\address{Faculty of Engineering and Architecture, Yeni Y\"{u}zy{\i}l University, Cevizliba\v{g}-Topkap{\i}, 34010  Istanbul, Turkey}
\date{\today}

\begin{abstract}

Rotating small AdS black holes exhibit the superradiant instability to low-frequency  scalar perturbations, which is amenable to a complete analytic description in four dimensions. In this paper, we extend this description to all higher dimensions, focusing on  slowly rotating charged AdS  black holes with a single angular momentum. We divide the spacetime of these black holes
into the near-horizon and far regions and  find  solutions to the scalar wave equation in each of these regions. Next, we perform the matching of these solutions in the overlap between the regions, by employing the idea that the orbital quantum number $ \ell  $ can be thought of as  an approximate integer. Thus, we obtain  the  complete low-frequency solution that allows us to calculate the complex frequency spectrum of quasinormal modes, whose imaginary part is determined by a small damping parameter. Finally, we find a remarkably instructive expression for the damping parameter, which appears to be a complex quantity in general. We show that the real part of the damping parameter can be used to give a  {\it universal} analytic  description of the superradiant instability for slowly rotating  charged AdS black holes in all spacetime dimensions.

\end{abstract}


\maketitle

\section{Introduction}

Superradiance is a property of rotating black holes by which waves of certain frequencies  are amplified when scattering by the black holes. The amplification effect occurs due to the extraction of rotational energy from the black holes. This is a wave analog of the hypothetical Penrose process \cite{penrose} wherein the extraction of  rotational energy occurs  by orbiting and breaking up a particle in the ergoregion of the black holes. Zel'dovich apparently first pointed out  that  a rotating  black hole  may possess the superradiant property just as an axially-symmetric body rotating  with an angular velocity $\Omega $ in a resonant cavity, where oscillation  modes  with the frequency $\omega $ and  azimuthal number $ m $,  fulfilling  the condition $\omega < m \Omega $, undergo amplification  \cite{zeldovich1,zeldovich2}. The superradiant property was independently discussed by Misner \cite{misner} as well, who demonstrated that scalar waves scattered by a Kerr black hole become amplified if the wave frequency $\omega < m \Omega_H $, where $ \Omega_H $ is the angular velocity of the black hole. Developing further Zel'dovich's heuristic idea on the possibility of  exponential amplification of waves, when surrounding a rotating black hole by a semitransparent mirror, Press and Teukolsky  \cite{press1} proposed  a ``black hole bomb" mechanism   as an intriguing probe of the strong superradiant instability in the black hole-mirror system.

Following the spirit of these fruitful ideas, Starobinsky \cite{starobinsky}  developed a quantitative theory of superradiance  for scalar waves in the Kerr metric. Similar  considerations  for electromagnetic and gravitational  waves  were  given in \cite{starchur}. Ever since, superradiance  and the black hole bomb mechanism  have been the subject of many investigations. The authors of works \cite{press2, press3} have shown that rotating black holes are stable against all modes of  massless bosonic  perturbations, whereas it appeared to be opposite for  massive bosonic perturbations \cite{damour,zouros, detweiler} (see also \cite{ag1} and references therein). It is the nonvanishing mass of a bosonic field that provides  a natural mirror around the black hole, thereby resulting in the superradiant  instability of  bound state modes of the bosonic field. In addition, it has recently been shown that the instability time scale for these bound state modes  may become orders of magnitude shorter in many cases of physical interest \cite{dolan, hod, roman1, cardoso1, witek, cardoso2} and thus giving rise to potentially observable effects of the superradiant instability. In this regard, the case of ultralight axions  appearing  in the ``axiverse" scenario  \cite{axiverse} of string theory compactifications  is  particularly  intriguing. It appears that for   axions  in  a certain mass range, the time scale of the superradiant instability becomes significantly short, creating gaps in the mass-spin spectrum of astrophysical black holes \cite{axiverse}.

Yet another physically more realistic setup for the black hole instability can occur in spacetimes with natural reflective boundaries. For example, the  ``confining-box" behavior of asymptotically anti-de Sitter (AdS) or G\"{o}del spacetimes can serve as a resonant cavity between the rotating black holes and spatial infinity. This in turn would result in the instability of superradiant modes of massless bosonic perturbations.  As shown in \cite{cardoso4}, small Kerr-AdS black holes do indeed exhibit the superradiant instability to  scalar perturbations (see also \cite{cardoso5, dias1, dias2, ran1} for related works). Using  the arguments of works \cite{cardoso4,  hreall}, a quantitative description of the superradiant instability of small rotating charged AdS black holes in five dimensions was given  in \cite{ad}. Meanwhile, the  unstable superradiant  modes of  scalar perturbations, which occur  for rotating black holes in the G\"{o}del universe of five-dimensional minimal ungauged supergravity,  were studied in \cite{roman2}  by using numerical calculations. More recently, the superradiant effects of a massive and charged scalar field  for small Reissner-Nordstr\"{o}m-AdS  black holes were analyzed  in all spacetime dimensions, by  employing  analytical and numerical techniques \cite{hwang}.
This analysis has  shown that  in odd  spacetime dimensions the matching procedure of \cite{ad} fails for some values of the orbital quantum number $ \ell $, thereby entailing the apparent absence of the  instability for these modes. This occurs because the near-horizon and far regions solutions to the Klein-Gordon equation have different functional dependence (involving the logarithmic term) that makes their matching impossible.  Solving numerically the Klein-Gordon equation for this special case,  where the analytical method is no longer valid, the authors of \cite{hwang}  have shown that the superradiant instability does exist  in this case as well.

In spite of the subtleties  with the matching procedure in  odd spacetime  dimensions,  one can still successfully use an analytical approach to give a complete  description of the superradiant instability in higher dimensions. In fact, this was demonstrated in \cite{aliev1} by  focusing on the black hole bomb model for scalar perturbations in five dimensions. The key ingredients of this description are based on a theoretical assumption that the orbital quantum number $ \ell $ is not exactly, but nearly integer. This allows one to perform the  matching of the near-horizon and far regions solutions  in their overlap region, avoiding the solution with a logarithmic term that excludes the matching procedure. The resulting  frequency spectrum of  bound state modes, which is calculated by taking  the corresponding limits as $ \ell $  approaches even or odd integers, shows that all $ \ell $  modes  undergo the superradiant instability.

The main aim of this paper is to continue the spirit of \cite{aliev1} and give  a complete analytic  description of  the superradiant  instability in all higher dimensions, focusing on the case of slowly rotating charged AdS  black holes with a single angular momentum. In Sec. II we begin  with the spacetime metric for a higher-dimensional Kerr-AdS black hole and  discuss  some of its basic properties. Assuming that the black hole may carry a small electric charge, we present the associated potential one-form  and calculate the electrostatic potential of the horizon. We also note that with a generic electric charge, the associated spacetime metric is obtained in the limit of slow rotation and by a simple rescaling of the mass parameter in the Kerr-AdS  metric under consideration. In Sec. III we examine a massless charged scalar field propagating in the  background of the higher-dimensional weakly charged Kerr-AdS black hole. Writing out the associated  Klein-Gordon equation, we show that this equation completely separates; it singles out the Laplace-Beltrami operator on a unite $ (N-3)$-sphere, where $ N $ is the number of spatial dimensions, and  yields two decoupled equations, the  radial and angular  equations. In Sec. IV we construct solutions the radial wave equation in the regime of low-frequency  perturbations: We divide the spacetime of  a small and slowly rotating charged AdS black hole into the near-horizon and far regions and  find solutions to the radial equation in each of these regions. Next, we perform the matching of these solutions in an intermediate region by utilizing the idea that the orbital quantum number $ \ell  $  can be thought of as  an approximate integer, as earlier suggested in \cite{aliev1}. This results in the  complete low-frequency solution to the radial wave equation. We also discuss the complex frequency spectrum of quasinormal modes, whose imaginary part is given by a small damping parameter. Finally, a remarkably instructive expression for the damping parameter is given in Sec. V, which turns out to be a complex quantity. We argue that the real part of this expression universally  describes the superradiant instability of slowly rotating  charged AdS black holes in all higher dimensions and  to all modes of scalar perturbations. In Sec. VI  we conclude with a  discussion of our results.

\section{The spacetime metric}

The spacetime  metric for general multiply rotating AdS black holes  in all higher dimensions was  found in \cite{glpp1}. We will focus on the higher-dimensional AdS black hole with a single angular momentum (the Kerr-AdS black hole), thus avoiding complications introduced by multiple rotations. The metric for this black hole is given by
 \begin{eqnarray}
ds^2 & = & -{{\Delta_r}\over {\Sigma}} \left(\,dt - \frac{a
\sin^2\theta}{\Xi}\,d\phi\,\right)^2 + {\Sigma \over~ \Delta_r}
dr^2 + {\Sigma \over ~\Delta_{\theta}}\,d\theta^{\,2} \nonumber
\\[3mm] &&
+ \,\frac{\Delta_{\theta}\sin^2\theta}{\Sigma} \left(a\, dt -
\frac{r^2+a^2}{\Xi} \,d\phi \right)^2 + r^2 \cos^2{\theta} \,
d\Omega_{N-3}^2\,,
\label{hdkads}
\end{eqnarray}
where $ N $ is the number of spatial dimensions ($ N\geq 3 $) and
\begin{equation}
d\Omega_{N-3}^2 =d {\chi_{1}}^2+ \sin^2{\chi_{1}}\,(\,d
{\chi_{2}}^2+\sin^2{\chi_{2}}\,(...d{\chi_{N-3}}^2...)\,)\,,
\label{sphmetric}
\end{equation}
is the metric on a unit $\,(N-3)$-sphere. The metric functions are given by
\begin{eqnarray}
\Delta_r &= &\left(r^2 + a^2\right)\left(1 +\frac{r^2}{l^2}\right)
- m \,r^{4-N} \,,~~~~\Sigma = r^2+ a^2 \,\cos^2\theta \,,
\nonumber \\[2mm]
\Delta_\theta & = & 1 -\frac{a^2}{l^2}
\,\cos^2\theta\,,~~~~~~ \Xi=1 -
\frac{a^2}{l^2}\,.
\label{metfunct}
\end{eqnarray}
Here $\, l\,$ denotes the curvature radius of the AdS space, which is given by the negative cosmological constant, $\,\Lambda=- l^{-2} N(N-1)/2 $.  The parameters $ m $  and  $ a $ are, respectively, determined  by the  mass  and angular momentum of the black hole,  as given in \cite{gpp1, dkt, aliev2, aliev3}. Meanwhile,  for the determinant of  metric (\ref{hdkads}) we have
\begin{equation}
\sqrt{-g}= \frac{\Sigma
\sin\theta}{\Xi}\,\sqrt{\gamma}\,r^{\,N-3}\,
\cos^{\,N-3}{\theta}\,,\label{determinant}
\end{equation}
where $\gamma$  is  the determinant of metric (\ref{sphmetric}).

Next, we need to know the angular velocity  and the electrostatic potential of the black hole, assuming that it  may possess a small electric charge. They can be obtained by using the time-translational and rotational isometries of  metric  (\ref{hdkads}), described  by two commuting Killing vectors  $ \xi_{(t)}=\partial_t $ and $  \xi_{(\phi)}=\partial_ \phi\,$. Defining a family of  locally nonrotating observers on orbits with constant $r$ and $\theta$, for which the  velocity vector  $ u^{\mu} $ obeys the condition $ u\cdot {\xi}_{(\phi)}=0 $, we find that the coordinate angular velocity of these observers  at the horizon  $\,r \rightarrow r_{+}\,(\Delta_r=0 ) $ is given by
\begin{equation}
\Omega_{H}= \frac{a\,\Xi}{r_{+}^2 + a^2}\,\,.
 \label{hvelocity}
\end{equation}
This in turn gives the angular velocity of the black hole.  With  this angular velocity, the co-rotating Killing vector  $ \chi = \xi_{(t)}+ \Omega_{H}\,\xi_{(\phi)} $  becomes  tangent to the null surface $ \Delta_r=0 $, thus correctly describing the isometry of the horizon (see \cite{aliev3} for details).

Meanwhile,  the potential one-form  for a small electric charge in the the background of metric (\ref{hdkads}) is determined by the difference between the timelike Killing isometries of this metric and those of its reference background, as shown in \cite{aliev2, aliev3}. Consequently, we have
\begin{equation}
A= -\frac{Q \, r^{4-N}}{(N-2)\,\Sigma}\left(dt- \frac{a
\sin^2\theta}{\Xi}\,d\phi \right),
 \label{hpotform}
\end{equation}
where the parameter $ Q $ is given by the electric charge of the black
hole through Gauss's law.  The associated contravariant components are given by
\begin{eqnarray}
A^{0}&=& \frac{Q\,r^{4-N}}{(N-2)\,\Sigma} \,\,\frac{r^2 + a^2}{\Delta_r}\,,~~~~~ A^{3} = \frac{Q\,r^{4-N}}{(N-2)\,\Sigma} \,\, \frac{a \Xi}{\Delta_r}\,,
\label{potcontras}
\end{eqnarray}
which, by equation (\ref{hpotform}), yield
\begin{equation}
A_{\mu} A^{\mu}= - \left(\frac{Q}{N-2}\right)^2\,\frac{r^{2(4-N)}}
{\Sigma \Delta_r}\,.
\label{sqpot}
\end{equation}
With these expressions in mind, it is now straightforward to show that the electrostatic potential of the horizon, defined as $ \,\Phi_H  = - A \cdot \chi \, $,  is given by
\begin{eqnarray}
\Phi_H   =  \frac{Q}{(N-2)}\,\,\frac{r_{+}^{4-N}}{r_{+}^2 + a^2}\,.
\label{hpot}
\end{eqnarray}
We also note  that by a rescaling  of the mass parameter in metric (\ref{hdkads}),
\begin{equation}
m\rightarrow m-q^2/r^{N-2}\,,
\label{resc}
\end{equation}
one can introduce a generic electric charge into the black hole spacetime. Then, for  $ N=3 $ we have the familiar Kerr-Newman-AdS metric in four dimensions,  in which case $ m=2 M $ and the potential one-form has the form as given in (\ref{hpotform}). However, in higher dimensions the system of Einsten-Maxwell equations becomes consistent only in the limit of slow rotation \cite{aliev4}. Thus, with equations (\ref{hpotform}) and (\ref{resc}) and keeping only linear in $a$  terms, the metric in (\ref{hdkads}) describes a slowly rotating higher-dimensional  AdS black hole  with an arbitrary amount of the electric charge. The horizon  of  such a black hole is governed by the equation $ \Delta_r=0  $ in (\ref{metfunct}), after performing the  rescaling  in accordance with (\ref{resc}) and dropping the $ a^2$ term. Thus, we have the equation
\begin{eqnarray}
r^{2(N-2)} \left(1+ \frac{r^2}{l^2}\right) -m r^{N-2} + q^2 = 0\,,
\label{metfunct1}
\end{eqnarray}
where the  parameter $ q $  is  related to the electric charge of the black hole by the relation
\begin{eqnarray}
q^2 &=&\frac{8 \pi G Q^2 }{(N-2)\,(N-1)}\,.
\label{physcharge1}
\end{eqnarray}
We note that in the limit of slow rotation, the  parameter  $ Q $  coincides with   electric charge of the AdS black hole  \cite{aliev3, aliev4}.

\section{Scalar Field}

We  now  consider  a massless scalar field $ \Phi $ propagating in the background of the Kerr-AdS black hole, given by  spacetime metric  (\ref{hdkads}). Assuming that the black hole may also have a small electric charge, it is straightforward to show that the Klein-Gordon equation
\begin{equation}
(\nabla_{\mu}- ie A_{\mu}) (\nabla^{\mu}- ie A^{\mu})
\Phi= 0\,,
\label{KGeq}
\end{equation}
where $ \nabla_{\mu} $  is a covariant derivative operator and $ e $ is the charge of the scalar field, can be written out in the form
\begin{eqnarray}
&&\frac{1}{\Sigma \, r^{N-3}}\frac{\partial}{\partial r} \left(\Delta_r\,  r^{N-3}\,
\frac{\partial \Phi}{\partial r}\right)
+\frac{1}{\Sigma \sin \theta \cos^{N-3} \theta}\,
\frac{\partial}{\partial\theta}\left( \Delta_{\theta} \sin \theta \cos^{N-3} \theta\,\frac{\partial \Phi}{\partial \theta}
\right)  \nonumber \\ [2mm]&&
+ g^{ab}\frac{\partial^2 \Phi}{\partial x^a {\partial x^a}}
- 2 i e A^a \frac{\partial \Phi}{\partial x^a}-e^2 A_a A^a \Phi
+\frac{1}{r^2 \cos^2\theta} \,\triangle_{(N-3)} \Phi
= 0\,.
\label{decomeq}
\end{eqnarray}
Here we have introduced the Laplace-Beltrami operator $ \triangle_{(N-3)}$ on a  unit $\,(N-3)$-sphere,
\begin{eqnarray}
\triangle_{(N-3)} \Phi &=&
\frac{1}{\sqrt{\gamma}}\, \frac{\partial}{\partial x^{\alpha}}\left(\sqrt{\gamma}\,\gamma^{\alpha \beta }\, \frac{\partial \Phi}{\partial x^{\beta}}\right),
\label{lboper}
\end{eqnarray}
decomposing the indices as $ \mu= \{r, \theta, \, a, \,\alpha \} $, where  $ a =0,3 \equiv t,\, \phi $ and $ \alpha = 1, ..., N-3 \,$.

Next, to separate variables in equation (\ref{decomeq}) we assume the ansatz  in the form
\begin{equation}
\Phi=e^{-i \omega t + i m \phi }\,  R(r)\, S(\theta)\, Y_j(\Omega)\,,
\label{sansatz}
\end{equation}
where $ m $ is the ``magnetic" quantum number that takes integer values being associated with rotation in the  $ \phi$-direction. In the following, we will focus on the case of  positive frequency $ (\omega > 0) $ and positive $ m $. The  hyperspherical harmonics $ Y_j(\Omega) $  are eigenfunctions of the Laplace-Beltrami operator (see e.g. \cite{atki}). The corresponding eigenvalues are given by
\begin{eqnarray}
\triangle_{(N-3)} Y_j(\Omega) &=& - j(j+N-4)  Y_j(\Omega).
\label{eigenv}
\end{eqnarray}
With this in mind, it is not difficult to show  that  separation ansatz  (\ref{sansatz}) yields  two decoupled ordinary differential equations; the angular equation
\begin{eqnarray}
\frac{1}{\sin \theta \cos^{N-3} \theta}
\frac{d}{d\theta}\left(\Delta_\theta \sin \theta \cos^{N-3} \theta \frac{d S}{d\theta}\right) +
\left[\lambda - \frac{1}{\Delta_\theta}\left(\frac{m \Xi}{\sin\theta} - a \omega \sin\theta\right)^2 \right.\nonumber \\[2mm]  \left.
- \frac{j(j+N-4)}{\cos^2\theta}\right] S &= & 0
\label{angular1}
\end{eqnarray}
and  the radial equation
\begin{equation}
\frac{\Delta_r}{r^{N-3}} \frac{d}{d r}\left(\Delta_r  r^{N-3}\,\frac{d R}{d r}\right) +U(r)\,R=0\,,
\label{radial1}
\end{equation}
where
\begin{eqnarray}
U(r)= -\Delta_r \left[\lambda  + \frac{j(j+N-4)a^2}{r^2}\right] + (r^2+a^2)^2 \left( \omega- \frac{a m \Xi}{r^2+a^2} - \frac{e Q}{N-2} \, \frac{r^{4-N}}{r^2+a^2}
\right)^2.
\label{radpot1}
\end{eqnarray}
It should be noted that angular equation (\ref{angular1}), subject to the regular boundary conditions at $ \theta=0 $ and  $ \theta=\pi/2 $, yields a well-defined eigenvalue problem for the separation constant $ \lambda = \lambda_{\ell}(\omega) $, where  $\ell $ is an integer which can interpreted as being an ``orbital" quantum number. The associated eigenfunctions are
AdS modified higher-dimensional spheroidal harmonics $ S(\theta)= S_{\ell\, m j}(\theta|a \omega) $. In general, the eigenvalues are  determined numerically, but in some special cases  they can also be calculated analytically, as shown in \cite{berti}. Similarly, assuming that   $a/l\ll 1$ and focusing on low frequencies  one can  show that
\begin{eqnarray}
\lambda &=& \ell(\ell+N-2) + \mathcal{O}\left(a^2 \omega^2, a^2/l^2 \right),
\label{eigenv1}
\end{eqnarray}
which implies  the  condition $ \ell \geq m + j $ as well.

To proceed further with radial equation (\ref{radial1}),  it is useful to introduce a new radial function, defined by
\begin{equation}
Y= \left[r^{N-3}(r^2 +a^2)\right]^{1/2}  R \,,
\label{newrad}
\end{equation}
and a new radial coordinate $ r_* $, \,given by
\begin{equation}
\frac{dr_*}{dr}=\frac{r^2+a^2}{\Delta_r}\,.
\label{tortoise}
\end{equation}
Using these relations in  equation (\ref{radial1}), we transform it to the Schr\"{o}dinger  form
\begin{eqnarray}
\frac{d^2 Y}{dr_*^2} +V(r) Y=0\,,
\label{radial2}
\end{eqnarray}
where the effective potential  is given by
\begin{eqnarray}
V(r)= \frac{U(r)}{(r^2+a^2)^2 }- \frac{1}{f} \frac{d^2 f}{dr_*^2}\,.
\label{effective}
\end{eqnarray}
Here $ U(r) $ is the same as that given in equation (\ref{radpot1}) and $ f = \left[r^{N-3}(r^2 +a^2)\right]^{1/2} $.

Next, we need to impose boundary conditions on  the propagating scalar waves at spatial infinity and at the horizon. Recalling that the AdS spacetime yields a natural reflective boundary  at spatial infinity due to its
confining-box behavior, it is tempting to impose the vanishing field boundary condition,
\begin{equation}
\Phi\rightarrow 0 \,,~~~~~~~ r\rightarrow \infty\, .
\label{boundinf}
\end{equation}
Meanwhile, from the physical point of view, it is  clear that at the horizon one must impose an ingoing wave boundary condition.  From equation (\ref{effective}) it follows that at the horizon,  $ r \rightarrow r_+ $, the effective potential  reduces to
\begin{eqnarray}
V(r_+ )&=&(\omega- m\Omega_{H} -e \Phi_{H})^2\,.
\label{p}
\end{eqnarray}
This  in turn yields the asymptotic solution that represents a purely ingoing wave at the horizon,
\begin{equation}
\Phi=e^{-i \omega t + i m  \phi}\, e^{-i(\omega- \omega_{p})r_*}S(\theta) Y_j(\Omega)\,,
\label{sansatzh}
\end{equation}
where $ \omega_{p} $ is  the threshold frequency, given by
\begin{equation}
\omega_{p}= m \Omega_{H} +e \Phi_{H}\,.
\label{bound}
\end{equation}
It  follows  that for the frequency range
\begin{equation}
0< \omega < \omega_{p}\,,
\label{fbound}
\end{equation}
the  phase velocity of the wave, $ v_{ph} = \omega/(\omega _{p}- \omega)$,  is in the opposite direction with respect to the group velocity, $ v_{gr}= -1 $. This fact signifies the appearance of superradiance, resulting in the energy outflow from the black hole.

It is clear that the above boundary conditions at spatial infinity and at the horizon render the frequency spectrum of bound state modes {\it quasinormal}, with complex frequencies. If the imaginary part of a characteristic frequency is positive, exponential growth  of the associated mode amplitude occurs, as follows from  decomposition (\ref{sansatz}). In this case, the system would eventually  develop instability. In what follows, we will describe this phenomenon for low-frequency modes in which case such a description appears to be amenable to analytic consideration.

\section{Low-frequency Solutions}

We will now construct  solutions to  radial wave equation (\ref{radial1}), focusing on low-frequency  perturbations, i.e. in the limit
when the frequency of the typical  perturbation is much less than the inverse of the horizon scale, $ \omega \ll  1/r_+\,$.  It is not difficult to show that even in this  case, one can not manage to find the general solution to equation (\ref{radial1}), by employing the methods which are known in the theory of  ordinary differential equations. On the other hand, this can be done  by dividing the spacetime into the near-horizon and far regions and looking  for solutions in each of these regions. Then the subsequent matching of these solutions in an overlapping region will yield  the complete low-frequency solution to equation (\ref{radial1}). The similar approach was first developed by  Starobinsky \cite{starobinsky} in the theory of superradiance  for a  Kerr black hole in four dimensions. To avoid potential complications, we will here  focus on a small  AdS black hole in the regime of slow rotation, i.e. when  $ r_+ \ll l $ and $ a \ll  r_+ $.  In the slow rotation regime, we will also  consider an arbitrary electric charge  for the black hole, using equation (\ref{resc}) in metric (\ref{hdkads}) and dropping all terms higher than linear order in rotation parameter $ a $. With these assumptions  in mind, we find solutions to radial equation (\ref{radial1}), first in the near-horizon region and then in the far-region of the spacetime.

\noindent
{\bf Near-horizon solution:\,} In the near-horizon region, we have  $ r- r_+ \ll 1/\omega $  and, for a slowly rotating AdS black hole of small-size, equation (\ref{radial1}) is approximated by
\begin{equation}
(N-2)^2\Delta_x \frac{d}{dx} \left(\Delta_x \frac{dR}{dx}\right) +\left[ x_+^{\frac{2(N-1)}{N-2}} \left(\omega-\omega_p\right)^2- \ell(\ell + N -2)\Delta_x \right] R=0\,,
 \label{nearrad1}
\end{equation}
where we have defined a new metric function  $ \Delta_x  $,  given by
\begin{equation}
\Delta_x = \Delta_r r^{2(N-3)} = (x-x_+)(x-x_-)\,,
\end{equation}
and  a new  coordinate  $ x=r^{N-2}$. Consequently, the quantities $ x_+ $ and  $ x_- $ correspond to  the radii of outer and inner horizons, which   are determined by equation (\ref{metfunct1}),  with $ r_+ \ll l $.  In obtaining equation (\ref{nearrad1}) we have also used equation (\ref{eigenv1}), dropping  small correction terms in it. Next, it is  not difficult to  show that  by  defining a new dimensionless  coordinate,
\begin{equation}
z=\frac{x-x_+}{x-x_-}\,\,,
\end{equation}
and performing  a few simple manipulations,  equation (\ref{nearrad1}) can be transformed into  the form
\begin{equation}
z (1-z) \frac{d^2R}{dz^2} +(1-z)\frac{d R}{dz} +\left[ \frac{1-z}{z}\,\Omega^2 -\frac{\ell}{N-2} \left(1+ \frac{\ell}{N-2}\right)\frac{1}{1-z}\right]R=0\,,
\label{nearrad2}
\end{equation}
where
\begin{equation}
\Omega=\frac{x_+^{\,\frac{N-1}{N-2}}}{N-2}\, \frac{\omega-\omega_p}{\,x_+ -x_-}\,.
\label{newsuperf}
\end{equation}
This is a hypergeometric  type  differential equation which can be solved  by the  substitution
\begin{equation}
R(z)=z^{i \Omega}\,(1-z)^{1+\frac{\ell}{N-2}}\,F(z),
\end{equation}
where the function $ F(z)= F(\alpha\,, \beta\,,\gamma, z) $ is given by the
standard hypergeometric equation
 \begin{eqnarray}
&& z(1-z)\frac{d^2 F}{dz^2} +\left[\gamma-(\alpha +\beta+1)\,z\right] \frac{d F}{dz} - \alpha \beta F = 0
\label{hyperg1}
\end{eqnarray}
with
\begin{equation}
\alpha= 1+\frac{\ell}{N-2} + 2 i \Omega \,,~~~~~~~\beta= 1+\frac{\ell}{N-2}\,,~~~~~~~
\gamma=1+2i\Omega\,.
\label{nearradpara}
\end{equation}
We are interested in the solution which at the horizon $ (z\rightarrow 0) $ obeys  the ingoing wave boundary condition  in (\ref {sansatzh}). It is not difficult to show that the desired solution has the form
\begin{equation}
R(z)= A_{(+)}^{in} \,z^{-i \Omega}\,(1-z)^{1+\frac{\ell}{N-2}}\,F\left(1+ \frac{\ell}{N-2}\,,~ 1+ \frac{\ell}{N-2} - 2 i \Omega\,,~ 1-2i\Omega\,,~ z\right),
\label{nearphys}
\end{equation}
where $ A_{(+)}^{in} $ is a constant. In the following we need to match  this solution to the far-region solution of equation (\ref{radial1}), in an intermediate region of their overlap. This in turn  requires us to know the behavior of this solution at large $ r $. However, it turns out that  the matching procedure fails for some values of  $ \ell $ (in odd spacetime dimensions) regardless of whether the black hole is static or rotating. For the static black hole, this subtlety was first noted in \cite{hwang}, emphasizing  the inevitability of a numerical  analysis in the case under consideration. Remarkably, one can  avoid the failure of the matching procedure by assuming that $ \ell $ is nearly integer, which can always be  thought of as a pure mathematical trick by introducing  a small deviation from its exact value. In the rotating case,  such an assumption could be argued physically as well, if one takes into account the correction terms in equation (\ref{eigenv1}). However,  we are dealing with the regime of slow rotation (ignoring  $a^2$ and higher-order terms) therefore, we will henceforth employ a formal mathematical trick, assuming that $ \ell $ is an approximate integer, which approaches the exact value when the small deviation from it vanishes.

Using now the standard   formula (see e.g. \cite{abramowitz}) that relates the hypergeometric functions of the arguments $ z $ and  $ 1- z $, we find that the large  $ r  \, (z \rightarrow 1) $ limit of solution (\ref{nearphys}) is given by
\begin{eqnarray}
R & \simeq & A_{(+)}^{in} \Gamma(1-2i\Omega)\left[\frac{\Gamma\left(-1- \frac{2\ell}{N-2} \right)\,(x_+ - x_-)^{1+\frac{\ell}{N-2}}}{\Gamma\left(-\frac{\ell}{N-2}\right)\,\Gamma\left(-\frac{\ell}{N-2} -2i\Omega\right)}\,\,r^{2 - N -\ell}
\right.\nonumber \\[2mm] && \left.
~~~~~~~~~~~~~~~~~~~~~~~~~~~~~~~~~~~~~ +\frac{\Gamma\left(1+\frac{2\ell}{N-2} \right)\,(x_+- x_-)^{-\frac{\ell}{N-2}}}
{\Gamma\left(1+\frac{\ell}{N-2}\right)\,\Gamma\left(1+\frac{\ell}{N-2} -2i\Omega\right)}
\, \,r^{\ell}\,\right].
\label{larnear}
\end{eqnarray}
We recall once again that the number $ \ell $  in this expression  is supposed to be nearly  integer since the quotient of gamma functions  $\Gamma\left(-1- \frac{2\ell}{N-2} \right)/\Gamma\left(-\frac{\ell}{N-2}\right)$ appearing
in the first line requires a special care for $ N \geq 4 $, yielding in some cases   divergent results for an exact integer $ \ell $.

\noindent
{\bf Far-region solution:\,}  In the far-horizon region, $ r- r_+ \gg  r_+ $ , and for the black hole under consideration it follows that  we can approximate  equation (\ref{radial1}) by its purely AdS  limit. Thus, we have the equation
\begin{equation}
\left(1+\frac{r^2}{l^2}\right) \frac{d^2R}{dr^2} +
\left[\frac{N-1}{r}+(N+1) \frac{r}{l^2}\right]\frac{dR}{dr}
+\left[\frac{\omega^2}{1+\frac{r^2}{l^2}}-\frac{\ell(\ell+N-2)}{r^2} \right]R=0\,,
\label{farr}
\end{equation}
where we again assume that  $ \ell $  is nearly integer.  Next, we define a new  radial coordinate $y$, given by
\begin{equation}
y= \left(1+\frac{r^2}{l^2}\right),
\end{equation}
in terms of which equation (\ref{farr}) takes the form
\begin{equation}
y(1-y)\frac{d^2R}{dy^2} +\left[1- \left(1+ \frac{N}{2}\right)y \right] \frac{dR}{dy} -\frac{1}{4}\,\left[\frac{\omega^2 l^2}{y}-\frac{\ell(\ell+ N-2)}{y-1}\right]R=0\,.
\label{adsrad1}
\end{equation}
The solution to this equation we are interested in must be regular  at the origin of the AdS space, $ r\rightarrow0 $, and satisfy the vanishing field condition at spatial infinity (see equation (\ref{boundinf})). It is straightforward to verify that  the substitution
\begin{equation}
R=y^{\frac{\omega l}{2}}(1-y)^{\frac{\ell}{2}}\, F(y)\,,
\end{equation}
where the hypergeometric function $ F(y) $  is given by equation (\ref{hyperg1}), with the associated  parameters
\begin{equation}
\alpha= \frac{N}{2}+\frac{\ell}{2} + \frac{\omega l}{2}\,,~~~~~~\beta= \frac{\ell}{2}  + \frac{\omega l}{2}\,,~~~~~~\gamma =1+ \omega l\,,
\end{equation}
results in the desired solution.  Finally, we find that
\begin{equation}
R(y)= A_\infty \, y^{-\frac{\ell}{2}- \frac{N}{2}}\,(1-y)^{\frac{\ell}{2}}\,F\left(\frac{N}{2}+\frac{\ell}{2} + \frac{\omega l}{2}\,,~~\frac{N}{2}+\frac{\ell}{2} -\frac{\omega l}{2}\,,~~ 1+ \frac{N}{2}\,, ~~1/y\right)\,,
\label{nearphys1}
\end{equation}
where $ A_\infty  $ is a constant. We also need to know the behavior of this solution at small distances, which becomes manifest when expressing it in terms of the hypergeometric functions of the argument $ 1-y $. After some algebra,  for  $ y\rightarrow 1 $  we find  the expansion
\begin{eqnarray}
&&R(r)\simeq  A_\infty \,(-1)^{\ell/2} \, \Gamma\left(1+ \frac{N}{2}\right)
\left[\frac{\Gamma\left(\frac{N}{2} + \ell-1\right)
\,l^{N+\ell-2} \,\,r^{2-N-\ell} }{\Gamma\left(\frac{N}{2}+\frac{\ell}{2} + \frac{\omega l}{2}\right) \Gamma\left(\frac{N}{2}+\frac{\ell}{2} - \frac{\omega l}{2}\right)}
\right. \nonumber\\[3mm]
&&
\left.
~~~~~~~~~~~~~~~~~~~~~~~~~~~~~~~~~~~~~~~~~~~~~~~
+\,\frac{\Gamma\left(1-\ell - \frac{N}{2}\right)\,l^{-\ell}\,\,r^{\ell}}
{\Gamma\left(1-\frac{\ell}{2} + \frac{\omega l}{2}\right) \Gamma\left(1-\frac{\ell}{2} - \frac{\omega l}{2}\right)}
\right],
\label{farradsmall}
\end{eqnarray}
which specifies  the small $ r $  behavior of  solution  (\ref{nearphys1}).  In order that this solution be finite at the origin of the AdS space, $ r=0 $, we must set the "quantization" condition
\begin{equation}
\frac{N}{2}+\frac{\ell}{2} - \frac{\omega l}{2} = -n ,
\label{quantization}
\end{equation}
which follows from the pole structure of the gamma function, $\Gamma\left(\frac{N}{2}+\frac{\ell}{2} - \frac{\omega l}{2}\right)=\infty $. Here  $ n $ is a non-negative integer which can be thought of as
a ``principal" quantum number. Clearly, this condition  governs the discrete  frequency spectrum for scalar perturbations in the  higher-dimensional AdS  spacetime, resulting in  the remarkably simple formula
\begin{equation}
\omega_n=\frac{2n+\ell+N}{l}\,.
\label{fspectrum}
\end{equation}
This is in agreement with the result  given in \cite{hwang}. Recall that we consider only positive frequency spectrum. Meanwhile, the presence of a black hole  would  render the decay of  bound state modes by tunneling the waves through the potential barrier into the horizon. In this case, as  mentioned above, the frequency spectrum becomes quasinormal, given  by the complex  frequencies
\begin{equation}
\omega= \omega_n + i\delta\,,
\label{complexf}
\end{equation}
where $ \delta $ is a small damping parameter that ``measures" the decay of bound state modes due to the presence of the black hole.

\noindent
{\bf Overlap region:\,} Comparing now equation (\ref{farradsmall}) with that given  in (\ref{larnear}), it is not difficult to see that  there exists an overlapping region, $ r_+ \ll r-r_+\ll 1/\omega $, of validity  for the near-horizon  and far-region solutions. Thus, with relations  (\ref{fspectrum}) and (\ref{complexf}) in mind, one can perform the matching of these solutions in the overlapping region. This  allows us to find the damping parameter by iteration and, to first order, it is given by
\begin{eqnarray}
\delta &=&  2 i\, \frac{(-1)^n}{n!}\,\,\frac{(x_+ - x_- )^{1+\frac{2\ell}{N-2}}}{l^{N + 2 \ell - 1}}\,\, \frac{\Gamma\left(1+\frac{\ell}{N-2}\right)}{\Gamma\left(1+\frac{2\ell}{N-2}\right)}\,\,
\frac{\Gamma\left(N+\ell+n\right)}{\Gamma\left(\frac{N}{2}+ n + 1\right) \Gamma\left(\frac{N}{2}+ \ell - 1 \right)}\times
\nonumber
\\[3mm] &&
\frac{\Gamma\left(-1-\frac{2\ell}{N-2}\right)}{\Gamma\left(-\frac{\ell}{N-2}\right)}\,\, \frac{\Gamma\left(1 + \frac{\ell}{N-2} -2i\Omega\right)}{\Gamma\left(-\frac{\ell}{N-2} -2i\Omega\right)}\,\, \frac{\Gamma\left(1- \ell- \frac{N}{2}\right)}{\Gamma\left(1- \ell- \frac{N}{2} -n \right)}\,.
\label{delta1}
\end{eqnarray}
Here the quantity $\Omega $  has the same form as  given in equation (\ref{newsuperf}), where one must now put  $\omega = \omega_n $.

\section{Instability}

As mentioned above, when the damping parameter is positive, a characteristic  field mode would exponentially grow its amplitude, resulting in the instability of the system. Therefore, to proceed further with expression (\ref{delta1}), we need to establish its sign for the cases of interest. We begin by simplifying  the quotients of gamma functions, appearing in the second line of this expression. Using the standard  functional relation for gamma functions $ \Gamma(z)\Gamma(1-z)=\pi/\sin{\pi z} $ and performing straightforward calculations, we find that
\begin{eqnarray}
\label{simrel1}
\frac{\Gamma\left(-1-\frac{2\ell}{N-2}\right)}{\Gamma\left(-\frac{\ell}{N-2}\right)}&=& - \frac{1}{2\cos\left(\frac{\pi \ell}{N-2}\right)}\,\,
\frac{\Gamma\left(1+\frac{\ell}{N-2}\right)}{\Gamma\left(2+ \frac{2\ell}{N-2}\right)}\,,\\[4mm]
\frac{\Gamma\left(1 + \frac{\ell}{N-2} -2i\Omega\right)}{\Gamma\left(-\frac{\ell}{N-2} -2i\Omega\right)}&=& -\frac{1}{\pi} \left|\Gamma\left(1 + \frac{\ell}{N-2} -2i\Omega\right)\right|^2\times
\nonumber
\\[4mm]&&
\left[\sin\left(\frac{\pi \ell}{N-2}\right)\cosh(2\pi \Omega) + i \cos\left(\frac{\pi \ell}{N-2}\right)\sinh(2\pi \Omega)\right],\\[4mm]
\frac{\Gamma\left(1- \ell- \frac{N}{2}\right)}{\Gamma\left(1- \ell- \frac{N}{2} -n \right)}&=& (-1)^n\,\prod_{k=1}^{n} \left(\frac{N}{2}+\ell +k -1\right).
\label{simrel3}
\end{eqnarray}
In obtaining equation (\ref{simrel3}) we have also used the relation $\Gamma(z+k)= (z)_k \,\Gamma(z) $, where $ (z)_k $ is the Pochhammer factorial, given by
\begin{equation}
(z)_k = z ( z+1) ...(z+k-1)= \prod_{i=1}^{k}(z+i-1)
\label{poch2}
\end{equation}
Substituting now these relations in (\ref{delta1}),  we find that the  expression for the damping parameter can be put in the following remarkable form
\begin{eqnarray}
\delta &=& \frac{(x_+ - x_- )^{1+\frac{2\ell}{N-2}}}{n! \,l^{N + 2 \ell - 1}}\,\, \frac{\Gamma^2\left(1+\frac{\ell}{N-2}\right)}{\Gamma\left(1+\frac{2\ell}{N-2}\right)\Gamma\left(2+\frac{2\ell}{N-2}\right)}\,\,
\frac{\left|\Gamma\left(1 + \frac{\ell}{N-2} -2i\Omega\right)\right|^2}{\pi \cos\left(\frac{\pi \ell}{N-2}\right)} \times
\nonumber
\\[3mm] &&
\left[i \sin\left(\frac{\pi \ell}{N-2}\right)\cosh(2\pi \Omega)
- \cos\left(\frac{\pi \ell}{N-2}\right)\sinh(2\pi \Omega)\right] \times
\nonumber
\\[3mm] &&
\frac{\Gamma\left(N+\ell+n\right)}{\Gamma\left(\frac{N}{2}+ n + 1\right) \Gamma\left(\frac{N}{2}+ \ell - 1 \right)} \,\prod_{k=1}^{n} \left(\frac{N}{2}+\ell +k -1\right).
\label{delta2}
\end{eqnarray}
We note that this is a complex  quantity, whose real part describes the damping of modes, while the imaginary part gives the frequency-shift of modes with respect to the AdS spectrum. It is easy to see that the overall sign of the real part is entirely determined by the sign of the quantity $ \Omega $ and it is positive for $ \Omega < 0$, i.e. in the superradiant regime. Meanwhile, no such sign-changing occurs for the imaginary part that behaves as  being  not sensitive to the superradiance.

In order to give further insight into expression (\ref{delta2}), let us now assume that the orbital quantum number $ \ell $ approaches a non-negative integer in the limit  $\epsilon \rightarrow 0 $ and consider the following exhaustive  cases: (i)  $ \frac{\ell}{N-2} = p + \epsilon $, where  $ p $ is a non-negative integer.\footnote{ Note that to retain the original form  of expression  (\ref{delta2}), for the sake of uniformity, we have not explicitly  used $ p $ therein, when discussing the above cases.} In this case, the imaginary part of (\ref{delta2}) vanishes  and the remaining  real part  correctly describes the instability of the associated modes in the superradiant regime, $ \Omega < 0 $. Clearly, this choice also encompasses  the case of superradiant instability for rotating AdS black holes in  four-dimensional spacetime $ (N=3) $, earlier considered in \cite{cardoso4}; (ii)  $ \frac{\ell}{N-2} \neq (p + 1/2) + \epsilon $, in which case, it is not difficult to see that the damping parameter in (\ref{delta2}) remains complex, describing both the  frequency-shift and superradiant instability of the associated modes, by its  imaginary and real parts, respectively; (iii)  $ \frac{\ell}{N-2} = (p + 1/2) + \epsilon $, which requires a special care and may occur only in odd spacetime dimensions. Substituting this in expression (\ref{delta2}) and  taking the limit  $\epsilon \rightarrow 0 $, we find that
\begin{eqnarray}
\delta &=& - \frac{(x_+ - x_- )^{1+\frac{2\ell}{N-2}}}{\pi\, n! \,l^{N + 2 \ell - 1}}\,\, \frac{\Gamma^2\left(1+\frac{\ell}{N-2}\right)\left|\Gamma\left(1 + \frac{\ell}{N-2} -2i\Omega\right)\right|^2}{\Gamma\left(1+\frac{2\ell}{N-2}\right)
\Gamma\left(2+\frac{2\ell}{N-2}\right)} \left[\sinh(2\pi \Omega) + \frac{i}{\pi \epsilon} \, \cosh(2\pi \Omega) \right]\times
\nonumber
\\[3mm] &&
\frac{\Gamma\left(N+\ell+n\right)}{\Gamma\left(\frac{N}{2}+ n + 1\right) \Gamma\left(\frac{N}{2}+ \ell - 1 \right)} \,\prod_{k=1}^{n} \left(\frac{N}{2}+\ell +k -1\right).
\label{delta3}
\end{eqnarray}
Again, we see that the real part of this expression becomes positive in the supperradiant regime, $ \Omega < 0 $, thus  resulting in exponential growth of the mode amplitudes. Meanwhile, its imaginary part contains $ 1/\epsilon\, $ type divergence  in the limit $ \epsilon \rightarrow 0 $. However, it should be remembered that we are dealing with small-size black holes: The horizon radius $ r_+ $ is small enough, fitting the range of validity of  the low-frequency solution under consideration, $ r_+ \ll r-r_+\ll 1/\omega $. With this in mind, one can argue that the ratio $ \frac{(x_+ - x_- )^{1+\frac{2\ell}{N-2}}}{\epsilon}\,$ showing up in the imaginary part of
(\ref{delta3}), to  high accuracy, can be driven to a small finite quantity  even  for the lowest mode, $p=0 $. This expectation is also confirmed by a numerical analysis of expression (\ref{delta3}). The  results of numerical calculations  are presented in Tables I-II.

\begin{table}
\label{5Dsugrastatic}
\caption{The damping parameter of fundamental modes  for a  Reissner-Nordstr\"{o}m-AdS  black hole  with $ r_+=0.01 $ in five and seven dimensions $(D=N+1); $  the scalar field charge $e=12$\,.}
\begin{ruledtabular}
\begin{tabular}{l l l }
$q/q_e $ &$\delta_{\frac{\ell}{N-2}= (p+1/2)+\epsilon}\,,\,~  N=4\,, ~\ell=1    \,,~\epsilon\rightarrow 10^{-8}\,; $  & $\delta_{\frac{\ell}{N-2}= (p+1/2)+\epsilon}\,,\,~N=6\,, ~\ell=2\,,  ~\epsilon\rightarrow  10^{-15}$ \\
\hline
0.1 & $-1.453\times 10^{-9} -0.370 \times 10^{-8}$\,i/$\epsilon $   &
$-1.509\times 10^{-16} -0.431 \times 10^{-15}$\,i/$\epsilon $\\
0.3 & $-6.342\times 10^{-10} -0.311 \times 10^{-8}$\,i/$\epsilon $ & $-1.013\times 10^{-16} -0.363 \times 10^{-15}$\,i/$\epsilon $    \\
0.5 & $+5.444\times 10^{-11} -0.210 \times 10^{-8}$\,i/$\epsilon $  & $-5.275\times 10^{-17} -0.246 \times 10^{-15}$\,i/$\epsilon $\\
0.7 & $+4.299 \times 10^{-10}-0.098\times 10^{-8}$\,i/$\epsilon $
  & $-1.497\times 10^{-17} -0.114 \times 10^{-15}$\,i/$\epsilon $\\
0.9 & $+3.174\times 10^{-10}-0.020 \times 10^{-8}$\,i/$\epsilon $     & $+ 2.209\times 10^{-18} -0.011 \times 10^{-15}$\,i/$\epsilon $  \\
\end{tabular}
\end{ruledtabular}
\end{table}

\begin{table}
\label{5Dsugrarotating}
\caption{ The damping parameter of fundamental modes  for a  Kerr-AdS  black hole  with $ r_+=0.01 $ in five and seven dimensions $(D=N+1); $
the magnetic quantum number $ m=1 \,.$ }
\begin{ruledtabular}
\begin{tabular}{l l l }
$ a/r_{+} $ & $\delta_{\frac{\ell}{N-2}= (p+1/2)+\epsilon}\,,\,~  N=4\,, ~\ell=1 \,,~\epsilon\rightarrow 10^{-8}\,; $  & $\delta_{\frac{\ell}{N-2}= (p+1/2)+\epsilon}\,,\,~N=6\,,~\ell=2\,,~\epsilon\rightarrow  10^{-15}$ \\
\hline
0.1 & $1.853\times 10^{-9} -0.378 \times 10^{-8}$\,i/$\epsilon $   &
$ 4.318\times 10^{-17} -0.437 \times 10^{-15}$\,i/$\epsilon $\\
0.2 & $5.639\times 10^{-9} -0.408 \times 10^{-8}$\,i/$\epsilon $ & $2.597\times 10^{-16} -0.443 \times 10^{-15}$\,i/$\epsilon $    \\
0.3 & $9.657\times 10^{-9} -0.468 \times 10^{-8}$\,i/$\epsilon $  & $4.790\times 10^{-16} -0.458 \times 10^{-15}$\,i/$\epsilon $\\
0.33 & $1.093 \times 10^{-8}-0.493\times 10^{-8}$\,i/$\epsilon $
  & $5.457\times 10^{-16} -0.465 \times 10^{-15}$\,i/$\epsilon $\\
\end{tabular}
\end{ruledtabular}
\end{table}
In Table I we give the numerical results  for  a small  Reissner-Nordstr\"{o}m-AdS   black hole in five  and  seven spacetime dimensions. We  normalize all quantities in terms of the AdS curvature scale $ l$  and  take $ l=1 $, for certainty. From equation  (\ref{metfunct1}), we find that the extreme charge of the black hole $ q_e = r_+^2 \,$. In both cases, the calculations are performed for $ r_+=0.01 $,~ $ e=12 \,$ and for the lowest modes ($ n=0,\, p=0 $). Accordingly, we take $ \ell=1 $ in five dimensions and  $ \ell=2 $ in seven dimensions.  We see that for sufficiently large values of the electric charge, the superradiant instability appears in both cases  when the real part of the damping parameter  becomes positive. We also see that  the imaginary part of this parameter can be thought of as representing a small frequency-shift in the spectrum by choosing   $\epsilon\rightarrow 10^{-8} $ in five dimensions and $\epsilon\rightarrow 10^{-15} $ in seven dimensions.
Table II presents the results of similar  numerical analysis of the damping parameter  for a  slowly rotating Kerr-AdS  black hole. We have
the superradiant  instability to all modes under consideration and again,  a small frequency-shift in the spectrum;  for  the $ l=1$  mode as  $\epsilon\rightarrow 10^{-8} $ and for  the $ l=2$ mode as  $\epsilon\rightarrow 10^{-15}\,. $  Altogether, the above numerical analysis shows that the  idea of an  approximate integer $ \ell $ works well and is sufficient to give an analytical description of the superradiant instability in odd spacetime dimensions for all $ \ell $, approaching the integer,  as given in (\ref{delta3}).

Finally, we conclude that the real part of expression (\ref{delta2}) can be regarded as giving  a {\it universal} description of the  superradiant instability for small and slowly rotating  charged AdS black holes in all spacetime dimensions. The analysis of expression (\ref{delta2}) shows that the instability time scale,  $\tau= 1/\delta $, significantly  grows  as  the number of dimensions increases.

\section{Conclusion}

The superradiant instability of small rotating  black holes  in four-dimensional spacetimes has been thoroughly investigated in the literature. Remarkably, for low-frequency  scalar perturbations this phenomenon appears to be amenable to a complete quantitative description for both black hole-mirror systems and AdS black holes. The basic idea of this description traces back to a pioneering work of Starobinsky \cite{starobinsky}, where the complete  low-frequency solution to the Klein-Gordon equation was found by matching  together the  near-horizon and far regions solutions.

However, it appears that there exist some subtleties with the matching procedure in higher dimensions, where it fails to be valid for certain modes of scalar perturbations. This makes the use of numerical integration inevitable \cite{hwang}, thereby creating a gap in the complete analytic description of the  superradiant instability in higher-dimensional spacetimes. In this paper, we have filled this gap, extending the complete analytic description of the black hole superradiant instability to all higher dimensions. As an instructive model of crucial importance, we have elaborated on the case of a small rotating charged AdS  black hole, in the regime of slow rotation and with a single angular momentum.

First, we have demonstrated that Klein-Gordon equation for a massless charged scalar field, propagating in the  background of a higher-dimensional weakly charged Kerr-AdS black hole completely separates. Then, focusing on the spacetime of a slowly rotating charged AdS black black hole, we have found solutions to the radial wave equation in the near-horizon and far regions of this spacetime. Utilizing the idea of our previous work \cite{aliev1}, which relies  on the assumption that the orbital quantum number $ \ell  $  can be considered as  an approximate integer, we have performed the matching of these solutions in the overlap region. Thus, we have obtained
the  complete low-frequency solution to the radial wave equation, resulting in the complex frequency spectrum for quasinormal scalar modes. The small damping parameter appearing in the imaginary part of this spectrum  measures the decay of bound state modes.

Finally, we have calculated the damping parameter, ending up with a manifestly instructive expression which appeared to be a complex  quantity. Performing a detailed analysis of this expression for all pertinent cases, as the orbital quantum number $ \ell  $ approaches a non-negative integer, we have concluded that  its real part correctly describes the negative damping of modes in the regime of superradiance, i.e. the  superradiant instability of the higher-dimensional AdS black holes under consideration.

{\it Note added.} While completing this paper, the work of \cite{dtur} appeared where the superradiant instability of slowly rotating (uncharged) AdS black holes in higher dimensions is also discussed, thoroughly utilizing the key idea of \cite{aliev1} on the ``nearly" integer  orbital quantum number.

\section{Acknowledgment}

The author is a member of the Science Academy, Turkey and thanks the Academy for stimulating encouragement.

\end{document}